\documentclass[aps,prl,showpacs,twocolumn]{revtex4}

\usepackage{graphicx}
\usepackage{dcolumn}
\begin{document}

\title{Measurement of the $6s - 7p$ transition probabilities in atomic 
cesium and a revised value for the weak charge $Q_W$}

\author{A.A. Vasilyev}
\author{I.M. Savukov}
\author{M.S. Safronova}
 \altaffiliation[Current address: ]{Electron and Optical Physics 
Division, National Institute of Standards and Technology, 
Gaithersburg, Maryland 20899-8410}
\author{H.G. Berry}
\affiliation{Department of Physics, 225 Nieuwland Science Hall\\
University of Notre Dame, Notre Dame, Indiana 46556}

\date{\today}

\begin{abstract}
  We have measured the $6s - 7p_{1/2,3/2}$ transition
     probabilities in atomic cesium using a direct
      absorption technique.  We use our result plus other
       previously measured transition rates to derive an
        accurate value of the vector transition polarizability $\beta$ and,
        consequently, re-evaluate the weak charge $Q_W$.
          Our derived value $Q_W=-72.65(49)$ agrees with the prediction
          of the standard  model to within one standard deviation.
\end{abstract}


\maketitle

Bennett and Wieman's measurement \cite{1} in 1999 of the ratio of 
the off-diagonal hyperfine amplitude $M_{\mathrm{hf}}$  to the vector  
polarizability $\beta$  for the $6s-7s$ transition in cesium enabled them 
to evaluate the weak charge $Q_W$  of the electroweak interaction.  
In this evaluation they used a theoretical value for $M_{\mathrm{hf}}$  
which has been verified in subsequent calculations \cite{2a,2b}. 
Their $Q_W$  value differs from the prediction of the standard 
model by almost 2.5 standard deviations and has stimulated 
several recent theoretical papers \cite{3,4,5,21} which calculate the 
parity-nonconserving transition amplitude $E_{\mathrm{PNC}}$ of the 
cesium $6s-7s$ transition. 

This recent interest suggested a careful study of all the measured 
parameters which go into such a test. The scalar and vector 
polarizabilities  $\alpha$ and $\beta$  can be calculated as sums 
involving the reduced matrix elements of the electric-dipole 
transition rates from the $6s$ and $7s$ states 
(Refs.~\cite{6,10a,11}): 

\begin{eqnarray}
\alpha&=&\frac{1}{6} \sum_n \left[ \rule[-2ex]{0em}{0ex} \langle 
7s 
\parallel D
\parallel np_{1/2} \rangle \langle np_{1/2}
\parallel D \parallel 6s \rangle \right. \nonumber\\&\times& \left( 
\frac{1}{E_{7s}-E_{np_{1/2}}}+ 
       \frac{1}{E_{6s}-E_{np_{1/2}}} \right)
  \nonumber  \\ \nonumber   \\&-&
 \langle 7s
\parallel D \parallel np_{3/2} \rangle \langle  
np_{3/2}  \parallel D \parallel 6s \rangle \nonumber \\ &\times& \left. \left( 
\frac{1}{E_{7s}-E_{np_{3/2}}}+ 
       \frac{1}{E_{6s}-E_{np_{3/2}}} \right)
 \right],   \label{eq1}
\end{eqnarray}
                     
\begin{eqnarray}
\beta&=&\frac{1}{6} \sum_n \left[ \rule[-2ex]{0em}{0ex} \langle 
7s 
\parallel D
\parallel np_{1/2} \rangle \langle np_{1/2}
\parallel D \parallel 6s \rangle \right. \nonumber\\&\times& \left( 
\frac{1}{E_{7s}-E_{np_{1/2}}}- 
       \frac{1}{E_{6s}-E_{np_{1/2}}} \right)
 \nonumber   \\ \nonumber   \\&+&\frac{1}{2}
 \langle 7s
\parallel D \parallel np_{3/2} \rangle \langle  
np_{3/2}  \parallel D \parallel 6s \rangle \nonumber \\ &\times& \left. \left( 
\frac{1}{E_{7s}-E_{np_{3/2}}}-
       \frac{1}{E_{6s}-E_{np_{3/2}}} \right)
 \right].   \label{eq2}
\end{eqnarray}

In Eqs.~(\ref{eq1},\ref{eq2}), dominant contributions come from the $n=6,7$ 
terms. Therefore, the most important contributions come from the 
$6s - 6p$, $7s - 6p$, $6s - 7p$, and $7p - 7s$ matrix elements. 
The dominant contribution to the uncertainties of  $\alpha$  and 
$\beta$  calculated using this direct summation method comes from 
the uncertainty of $6s - 7p_{3/2}$ matrix element \cite{11}. 

In this paper, we present new measurements of $6s-7p$ transition 
rates.  The sum needed for the vector polarizability has some 
severe cancellations; hence, we use the experimentally 
well-determined $\alpha/\beta$ ratio   \cite{7} and our new 
measurement to determine $\beta$, and, consequently,  re-evaluate 
the  weak charge $Q_W$.

The best previous measurement of the $6s-7p$ transition rates was 
a photographic optical absorption measurement utilizing the hook 
method \cite{8}.  The relative measurement relied on the known 
values for the  $6s-6p_{3/2}$ transition.  In order to 
measure the transition probability directly, we have made an 
absolute absorption measurement of  laser light passing through a 
known number of cesium atoms for each transition. 

An electrically heated and insulated cesium cell, at temperatures 
between room temperature and 90~$^{\circ}$C provided a 5~cm long 
target for laser light close to the resonant wavelengths of 
  455~nm ($7p_{3/2}$) and 459~nm ($7p_{1/2}$). The cell temperature was 
measured with a multiprobe NIST calibrated K-type thermocouple 
thermometer with an accuracy of 0.1~$^\circ$C degree.  The needed 
blue light is produced by direct second harmonic generation (SHG) 
from a potassium niobate (KNbO$_3$) $5\times5\times5$~mm crystal 
pumped with a Coherent Model MBR-110 Titanium:Sapphire (Ti:Sapphire) ring 
laser. 

 To acquire each single absorption spectrum we scan the 
Ti:Sapphire laser over a  frequency range of 15~GHz during a 50 
seconds time interval.  We observe two well-resolved absorption 
peaks during each scan, since the separation between the hyperfine 
states of the $6s$ ground state  equals 9.19~GHz. 
The hyperfine structure of the $7p$ level is not resolved due to 
the much larger Doppler broadening of approximately
 750~MHz at 65~$^{\circ}$C. 
  Further 
experimental details will be published later \cite{9}.  

\begin{figure}  
\includegraphics[scale=0.45]{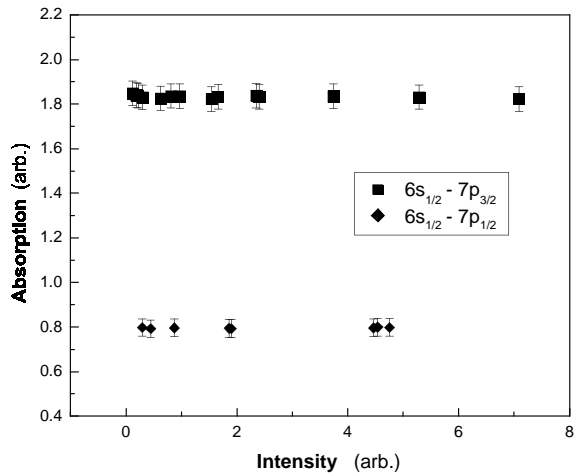}   
\caption{\label{fig1} Comparison of absorption rates for several 
laser intensities for the $6s-7p$ transitions at a fixed cesium 
cell temperature.}                            
\end{figure}

Since saturation might significantly influence the absorption measurements,
  we have measured the absorption for several different laser intensities,
   as shown in Figure~\ref{fig1}.  Similar results are also obtained for several
    different cell temperatures.  We have made an average of
     seventy sets of data to obtain the transition probabilities
      for the $6s - 7p_{3/2}$ transition of $1.836(18)\times10^6$~s$^{-1}$ and
       for the $6s - 7p_{1/2}$ transition of $7.934(80)\times10^5$~$s^{-1}$.
        Adding other uncertainties (due to temperature measurements and
         Cs vapor pressure) to these values yields results accurate to 
         1.6\%. The corresponding 
          reduced matrix elements are compared with theory \cite{11,14,15} and
           experiment \cite{8} in Table~\ref{tab1}.
           
First, we calculate the value of  $\alpha$  using
 the formula of Eq.~(\ref{eq1}). In Table~\ref{tab2},
  we list the values of electric-dipole matrix 
elements used in this calculation (present, Refs.~\cite{11,12,16}) 
together with  the uncertainty of each matrix element and its 
contribution to the uncertainty in the value of $\alpha$. We also 
list the contributions to the value of  $\alpha$ from the terms 
with $n > 7$ and from the term  $\alpha_{vc}$ which compensates 
for the excitations from the core to the valence shell violating 
Pauli principle; these very small contributions are taken from 
Ref.~\cite{11}.  As we see from Table~\ref{tab2}, the uncertainty 
in  $\alpha$ is dominated by the uncertainty in the value of the 
$7p_{3/2} - 6s$ matrix element. Therefore, our more accurate 
measurement of the $7p_{3/2} - 6s$ matrix element allows a 
significant decrease of the uncertainty in the value of   $\alpha$ 
(and correspondingly $\beta$) obtained by this method. 
          
In more detail, dominant contributions to the scalar and vector
 polarizabilities $\alpha$  and $\beta$ come from matrix elements 
 of terms with $n = 6$ and $n = 7$, while $n = 8$ and $n = 9$ contributions
  are relatively small but significant. The contributions from
   the terms with $n > 9$ are very small (less than 0.4\% according to \cite{11}). 
   Therefore, the values of only eight matrix elements are needed to 
   be known to high accuracy to produce accurate values of $\alpha$  and $\beta$:
    $6p_{1/2} - 6s$, $6p_{3/2} - 6s$, $7p_{1/2} - 6s$, $7p_{3/2} - 6s$,
     $6p_{1/2} - 7s$, $6p_{3/2} - 7s$, $7p_{1/2} - 7s$, and $7p_{3/2} - 7s$.
      The values of the $6p - 6s$ matrix elements were measured
       to better than 0.15\% precision in \cite{12}. 
       The values of the $7p - 7s$ matrix elements were derived in Ref.~\cite{11}
        from the experimental value of the Stark shift from Ref.~\cite{23} 
        with 0.15\% precision. These experimental values
       are in excellent agreement  with all high-precision theoretical
        calculations \cite{11,14,15}.
           The electric-dipole matrix elements for $7p_{1/2} - 6s$ and $7p_{3/2} - 6s$
            transitions are measured in this work with 0.8\% accuracy.
             The previous measurement of the $7p_{3/2} - 6s$ matrix element from \cite{8} 
             has 1.7\% uncertainty which gave the dominant contribution
              to the uncertainties of the recommended values of $\alpha$  and $\beta$
                in Ref.~\cite{11}. The $7p_{1/2} - 6s$ and $7p_{3/2} - 6s$
                 matrix elements are also difficult to  calculate accurately
                  (see, for example, Ref.~\cite{11} for discussion).
   The matrix elements for the $7s - 6p$ transitions are derived from 
   the measurement of the $7s$ lifetime conducted in Ref.~\cite{16}.
    The ratio of the reduced matrix elements for $7s - 6p_{3/2}$ and $7s - 6p_{1/2}$
     transitions is taken to be $R = 1.528 \pm 0.004$ based on 
      theoretical calculations \cite{11,14,15}.
       The uncertainty of the ratio does not significantly affect
        the uncertainties of the reduced matrix elements.
        We used the theoretical values for matrix elements with $n=8, 9$
        (the values of $6s-8p$ matrix elements are 
          taken from Ref.~\cite{11}) and the 
experimental values of energies from \cite{10} in evaluating 
formula of Eq.~(\ref{eq1}). We obtain the final value for 
           the scalar transition polarizability   $\alpha=269.7(1.1)$~a.u..
            Table~\ref{tab2} shows that 98.5\% of this value 
              comes from the experimentally derived contributions ($n=6,7$).

  \begin{table} [b]
\caption{\label{tab1} A comparison of theoretical and experimental 
reduced electric-dipole matrix elements (a.u.) for $7p_{1/2} - 6s$ 
and $7p_{3/2} - 6s$ transitions in cesium. In 
Ref.~\protect\cite{8}, $6s - 7p$ oscillator strength were 
normalized to the value of $6s - 6p_{3/2}$ oscillator strength. We 
have re-normalized those values to the most recently measured 
value of $6s - 6p_{3/2}$ oscillator strength from 
Ref.~\protect\cite{12}. } 
\begin{ruledtabular}
\begin{tabular}{ldd} 
\multicolumn{1}{c}{Ref.} &
\multicolumn{1}{c}{$7p_{1/2} - 6s$} & 
\multicolumn{1}{c}{$7p_{3/2} - 6s$} \\
      \hline 
Theory~\protect\cite{11}  &  0.279 &  0.576\\ 
 Theory~\protect\cite{14} &   0.275 &  0.583\\
Theory~\protect\cite{15}  &  0.280  & 0.576\\ 
  Expt.~\protect\cite{8} & 0.2825(20) & 0.5795(100)\\
   This work  &0.2757(20) & 0.5856(50)\\    
\end{tabular}
\end{ruledtabular}
\end{table}         

\begin{figure} 
\includegraphics[scale=0.47]{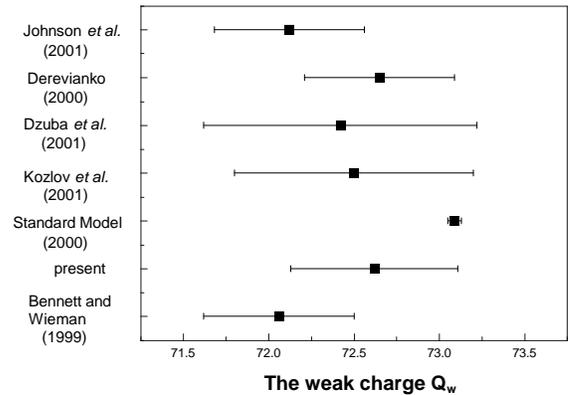}   
\caption{ \label{fig2} Comparison of recent theoretical and 
experimental determinations of $Q_W$.}             
\end{figure}

The parity-nonconserving effects in cesium give rise to a non-zero 
amplitude $E_{\mathrm{PNC}}$ for the $6s-7s$ transition forbidden by 
parity-selection rules. In Ref.~\cite{17}, the nuclear 
spin-independent average  ${\mathrm{Im}}(E_{\mathrm{PNC}})/\beta$ was 
measured to be $-1.5935(56)$~mV/cm. This value was combined in 
Ref.~\cite{1} with a measurement of  $\beta= 27.024(43)_{\mathrm{expt}}
(67)_{\mathrm{theor}}~a^3_0$,  conducted in the same work, and 
with an average of theoretical calculations \cite{6,18}  $ 
E_{\mathrm{PNC}} = 0.9065(36)\times10^{-11} i e a_0 Q_W/N,  
 $ where $N$ is the number of neutrons and $a_0$ is the Bohr radius,
to give the value of the weak charge $Q_W$.
 We note that the accuracy of the theoretical calculation of $E_{\rm PNC}$
  was taken in Ref.~\cite{1} to be 0.4\% based on the comparison
   of the  theoretical calculations of various atomic properties 
   conducted by the authors of Refs.~\cite{6,18} with experiment. 
     The resulting value of the weak charge 
  $  Q_W=-72.06(28)_{\mathrm{expt}}(34)_{\mathrm{theor}}
 $          \cite{1} 
was found to differ from  the value predicted by the standard  
model 
 $
  Q_W^{\mathrm{SM}}=-73.09(3)
$ from \cite{19} by $2.3\sigma$.
 Using our experimental result and the 
analysis given above to determine $\alpha$, plus the measurement 
by Cho {\it et al} \cite{7} of the  $\alpha/\beta$ ratio, we 
derive the almost completely experimentally determined  value  
$\beta=27.22(11)~a_0^3$. We use this result to determine the value 
of weak charge 
 $
    Q_W=-72.58(49),       
  $
which differs by only $1.1\sigma$ from the one  predicted by the 
standard model \cite{19}. However, the theoretical calculations of 
$E_{\mathrm{PNC}}$ have been improved recently to include Breit and 
vacuum-polarization corrections to the PNC amplitude 
\cite{3,4,5,21}. The revised value of the PNC amplitude, given in 
Ref.~\cite{5}, which is the average of three most accurate 
calculations \cite{6,18,4} and includes contributions from Breit 
and vacuum-polarization corrections \cite{3,5} is 
 $
     E_{\mathrm{PNC}} = 0.9057(37)\times10^{-11} i e a_0 \,Q_W/N.   
 $        
Combining this value with experiment \cite{17} and our result for 
$\beta$  we obtain our final value for the weak charge 
  $$
    Q_W=-72.65(49),
  $$
which is in agreement with the value predicted by 
the standard model \cite{19} to $1\sigma$.  In Figure~\ref{fig2}, we compare 
these results with recent calculations for the weak charge $Q_W$. 
                       
In conclusion, we have measured the probabilities of the $6s 
- 7p_{1/2,3/2}$ transitions in atomic cesium using a direct 
absorption technique.  We then indicate a  straightforward method 
to determine the scalar transition polarizability $\alpha$ based 
almost completely on experimentally determined atomic parameters. 
Including a previous accurate experimental determination of the 
 $\alpha/\beta$ ratio yields a value for the vector polarizability 
$\beta$ and for the weak charge $Q_W$.  

Our derived value for $Q_W$ agrees with the  value predicted by 
the standard model to within one standard deviation.  We compared 
the result with that of Bennett and Wieman \cite{1} and also with recent 
atomic calculations \cite{3,4,5,21} in Figure~\ref{fig2}.
  Future improvements in 
this method of estimating $Q_W$ can come from better calculations 
(or experimental measurements) of the $7s-6p$  transition rates 
and also from Tanner's measurement in progress (private 
communication) of the ratio of the $6s-7p$ transition rates.  
                                 
\begin{acknowledgments}
 We are grateful for financial support of this research from the 
Research Corporation, grant number Research Opportunity Award 
RA0243, and from the University of Notre Dame Graduate School.  We 
also acknowledge helpful discussions with Professor Walter 
Johnson and experimental support and discussions with Professor 
Carol Tanner. 
\end{acknowledgments}      
\begin{table*} 
\caption{\label{tab2} Contributions to the scalar $6s-7s$ 
transition polarizability   $\alpha$ in Cs and their uncertainties 
in a.u.}  \begin{ruledtabular} 
\begin{tabular} {lddddddd}
 \multicolumn{1}{l}{$n$} & \multicolumn{1}{c}{$d$} & 
\multicolumn{1}{c}{$\delta d(\%)$} & \multicolumn{1}{c}{$\delta 
\alpha$} & \multicolumn{1}{c}{$d$} & \multicolumn{1}{c}{$\delta 
d(\%)$} & \multicolumn{1}{c}{$\delta \alpha$} & 
\multicolumn{1}{c}{$\alpha$} \\ \hline \multicolumn{1}{l}{} & 
\multicolumn{3}{c}{$7s-6p_{1/2}$}  & 
\multicolumn{3}{c}{$np_{1/2}-6s$} & \multicolumn{1}{l}{} \\ \hline 
    6 & -4.236\footnotemark[1] & 0.5 &   0.16 & -4.489\footnotemark[2] & 0.1  &  0.05  & 32.32\\
    7 & 10.308\footnotemark[3] & 0.1 &   0.06 & -0.276\footnotemark[4] & 0.8  &  0.28  & 36.97\\
    8 & -0.915 & 2.0 &   0.01 &  0.081\footnotemark[3] & 3.0  &  0.01  &  0.48\\
    9 & -0.347 & 6.0 &   0.00 &  0.043 &10.0  &  0.01  &  0.08\\
\hline \multicolumn{1}{l}{} & \multicolumn{3}{c}{$7s-6p_{3/2}$} & 
\multicolumn{3}{c}{$np_{3/2}-6s$} & \multicolumn{1}{l}{} \\ \hline 
    6 &  6.473\footnotemark[1] & 0.5  &  0.46 & -6.324\footnotemark[2] & 0.1 &   0.11&   92.47\\
    7 &-14.320\footnotemark[3] & 0.1  &  0.16 & -0.586\footnotemark[4] & 0.8 &   0.78&  103.90\\
    8 &  1.622 & 2.0  &  0.05 &  0.218\footnotemark[3] & 3.0 &   0.07&    2.28\\
    9 &  0.678 & 6.0  &  0.03 &  0.127 &10.0 &   0.05&    0.46\\
\hline Expt. $\alpha_{6,7}$&&&&&&&265.66(98)\\   
 Theor. $\alpha_{8,9}$&&&&&&&3.3(1)\\  
 Theor. $\alpha_{>9}$     
&&&&&&&0.90(45)\footnotemark[3]\\ Theor. $\alpha_{vc}$   
&&&&&&&-0.2(1)\footnotemark[3]\\ 
$\alpha_{\mathrm{total}}$&&&&&&&269.7(1.1)\\ 
   \end{tabular}                   
\end{ruledtabular} 
\footnotetext[1]{Ref.~\protect\cite{16}}  
\footnotetext[2]{Ref.~\protect\cite{12}}   
\footnotetext[3]{Ref.~\protect\cite{11}} 
 \footnotetext[4]{This 
work} 
\end{table*}

\end{document}